\UseRawInputEncoding

\documentclass[journal]{IEEEtran}
\usepackage{cite}
\usepackage{amsmath,amssymb,amsfonts}
\usepackage{algorithm}
\usepackage{graphicx}
\usepackage{textcomp}
\usepackage{xcolor}
\usepackage{subfigure}
\usepackage{url}
\usepackage{amsmath}
\usepackage{graphicx}
\usepackage{makecell}

\ifCLASSINFOpdf
\else
\fi

\begin{document}
%
\title{Data-Driven Compressed Sensing for \\ Massive Wireless Access}

\author{Yanna~Bai,
        Wei~Chen,~\IEEEmembership{Senior~Member,~IEEE,}
        Feifei~Sun,
        Bo~Ai,~\IEEEmembership{Fellow,~IEEE},
        Petar~Popovski,~\IEEEmembership{Fellow,~IEEE}
\thanks{The manuscript has been accepted by IEEE Communication Magazine at September 16, 2022.}
\thanks{Yanna Bai and Wei Chen are with the State Key Laboratory of Rail Traffic Control and Safety, Beijing Jiaotong University, Beijing 100044, China, and also with Frontiers Science Center for Smart High-speed Railway System, Beijing 100044, China. (Email: yannabai, weich@bjtu.edu.cn).}
\thanks{ Feifei Sun is with Samsung Research Institute China –Beijing (SRC-B) Beijing, P.R.China (Email: feifei.sun@samsung.com).}
\thanks{Bo Ai is with the State Key Laboratory of Rail Traffic Control and Safety, Beijing Jiaotong University, Beijing 100044, China, and also with Research Center of Networks and Communications, Peng Cheng Laboratory, Shenzhen 518055, China, and also with Henan Joint International Research Laboratory of Intelligent Networking and Data Analysis, Zhengzhou University, Zhengzhou 450001, China. (Email: boai@bjtu.edu.cn).}
\thanks{Petar Popovski is with Aalborg University, Denmark.}
\thanks{Corresponding author: Wei Chen.}}

\maketitle

\begin{abstract}
The central challenge in massive machine-type communications (mMTC) is to connect a large number of uncoordinated devices through a limited spectrum. The typical mMTC communication pattern is sporadic, with short packets. This could be exploited in grant-free random access in which the activity detection, channel estimation, and data recovery are formulated as a sparse recovery problem and solved via compressed sensing algorithms. This approach results in new challenges in terms of high computational complexity and latency. We present how data-driven methods can be applied in grant-free random access and demonstrate the performance gains. Variations of neural networks for the problem are discussed, as well as future challenges and potential directions.

\end{abstract}


%
\IEEEpeerreviewmaketitle

\section{Introduction}

\IEEEPARstart{M}{assive} machine-type communication (mMTC) is one of three core services of 5G that characterized by sporadic, uplink communication, and low rates with small data packets. Already in 5G, it was considered that the grant-based access channel procedures of LTE need to be refined for mMTC. In grant-based access, active users send a randomly selected orthogonal preamble to the base station (BS) and wait for the responses. The users that received a response send connection requests for uplink resources. If the BS successfully decodes the contention request, it sends a contention-resolution message. Otherwise, the users have to repeat the hand-shaking procedure. The limited number of preambles is a bottleneck upon massive, batch arrivals of packets, and complicated signaling exchanges result in high overhead/inefficiency for small data packets. Grant-free random access has been widely seen as an approach that can overcome these drawbacks~\cite{8454392}.

\par
In grant-free random access, active users directly transmit the predefined preambles and data to the BS without waiting for the grant from the BS. This greatly reduces the signaling overhead and access delay, but brings additional uncertainty and complexity due to the unknown user activity. By exploring the natural sparsity of sporadic communication in mMTC, compressed sensing (CS)~\cite{1614066} is applied in grant-free random access~\cite{8454392}, where the user activity detection and channel estimation can be addressed jointly by solving a sparse recovery problem.

\begin{figure*}[t]%
\centering%
\includegraphics[width=0.95\textwidth]{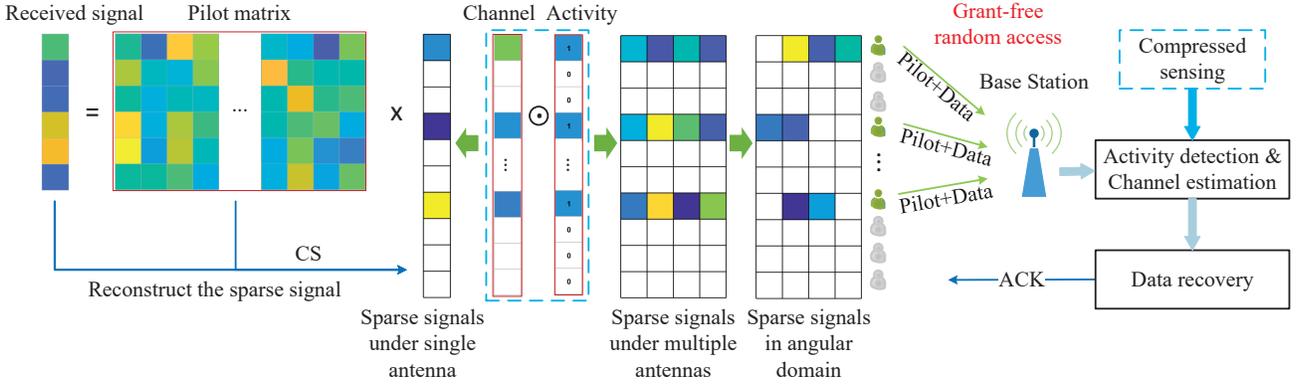}%
\DeclareGraphicsExtensions. \caption{An illustration of  CS-based grant-free random access scheme. The channel signal of all users can be thought as a sparse signal due to sporadic communication. Different rows of the sparse signal correspond to different users. Different colors of the spare signal represent channel coefficients, and blank represents zero. } \label{fig:CS-RA}
\end{figure*}%
\par
In recent years, applying data-driven methods to solve the problem in massive wireless access become a promising direction. For example, in data-driven methods, neural networks (NNs) can learn the hard-to-model activation patterns of devices from the training data, such that it can predict frequently activated devices more accurately. In addition, data-driven methods achieve a significant reduction in computational complexity, which are important to latency-sensitive applications.

\par
In this paper, we review the data-driven CS methods for user activity detection and channel estimation in grant-free access for mMTC. Firstly, we briefly introduce the CS-based massive access and traditional model-driven optimization methods. Then, we investigate various aspects of data-driven CS methods for massive access. Lastly, we analyze the NN design problem, which can provide a guide for designing customized NN in practical scenarios. In addition, we discuss main challenges in data-driven CS, present potential solutions, and indicate future directions.

\section{CS-based Grant-free Random Access}

\par
In grant-free access, each user is assigned a unique preamble, also referred to as the pilot sequence~\cite{8454392}. As shown in Fig.~\ref{fig:CS-RA}, in each coherence time, active users directly transmit their pilots and data to the BS, and the BS first detects the active users according to their pilots and estimates their channel, and then decodes their data. As the data transmission mostly is quasi-periodic or event-driven, only a small subset of devices is active at a given access instant. Thus, the activity of all users is a sparse signal and the user activity detection leads naturally to solving a CS problem. With known pilots of all users, the BS performs joint activity detection and channel estimation (JADCE) by recovering a sparse signal. However, it is very expensive to assign an orthogonal pilot for each user due to the limited spectrum resource and massive users. Thus, non-orthogonal pilot sequences are needed to support a vast number of users with limited pilot resources. This renders the massive access problem underdetermined and equivalent to the classical CS problem.


\par
Model-driven methods for this CS-based massive access problem explicitly utilize the characteristics, e.g., sparsity, to design customized optimization algorithms. However, those methods are usually computationally expensive and have slow convergence. As a viable alternative, data-driven methods learn from data to solve the detection/estimation problem, which offers improved performance and reduced computational complexity.

Data-driven CS methods use NNs to solve the sparse recovery problem for activity detection or/and channel estimation. While model-driven methods rely on mathematical models to the formulation of each iterative step, data-driven methods learns the parameters of NN from available training data. In the training phase of data-driven methods, the network adjusts its parameters to minimize the loss function, e.g., the mean square error between the estimated user activity vector and the ground truth. When inference is carried out, the received pilot signal at the BS is fed into the well-trained network, and the output of the network is the detected active user vector or estimated channel. As the pilot sequences are known, plenty of training data via simulators can be generated. Furthermore, real data can be used for training if available, which is more beneficial as the NN may learn more features from the real data, such as user activity patterns.

Data-driven methods for CS have been widely studied in literature. As confirmed in~\cite{gregor2010learning}, the trainable version of Li and Osher¡¯s coordinate descent method has the potential to achieve the target reconstruction performance 10 times faster than the original algorithm. This provides a considerable gain in access latency for time-sensitive applications. Data-driven methods also improve the CS estimation accuracy as observed in~\cite{gregor2010learning,7447163,7934066,9601284,9432908,9634107,9605579,9390399,9252937,9484069,9508782,9174792,9500572}. The performance gain of data-driven methods arises from several shortcomings of traditional CS methods, including the model error due to imperfect modeling, the structure error due to relaxation of the original problem, and the convergence error due to local optima.

\begin{table*}[]
\centering  
\setlength{\abovecaptionskip}{0pt}%
\setlength{\belowcaptionskip}{6pt}%
\caption{The classification of mentioned data driven methods.}\label{networks}
\resizebox{\textwidth}{!}{
\begin{tabular}{|c|c|c|c|c|}
\hline
category &Ref. &Network &Input/output &Description   \\ \hline\hline

CS +Data-Driven &\cite{gregor2010learning}  &Fully connected network (FNN) &The measurement/sparse signal  &The network is obtained via unfolding the iterative soft-thresholding algorithm (ISTA), named learned ISTA (LISTA).  \\\hline

CS +Data-Driven &\cite{7447163}  &Stacked denoising autoencoder &The sparse signal/sparse signal  &The network can be used for linear and mildly nonlinear measurements. \\\hline

CS +Data-Driven &\cite{7934066}  &FNN &The measurement/sparse signal  &The network is obtained via unfolding approximate message passing (AMP), named learned AMP (LAMP).  \\\hline

CS +Data-Driven &\cite{9601284}  &FNN &The measurement and measurement matrix/sparse signal  &The network is adaptive to varying model scenarios by combining the measurement matrix into network. \\\hline

CS +Data-Driven &\cite{9500572}  &Generative network &The measurement/sparse signal  &The network can quickly adapt to new measurement matrix in several iterations.  \\ \hline

CS + Data-Driven + Massive access &\cite{9432908}  &FNN &The measurement/non-zero indices  &Authors explores the low-rank property in massive MIMO and utilize it in network design to reduce the number of network parameters. \\\hline

CS + Data-Driven + Massive access &\cite{9634107}  &Long short-term memory (LSTM) &The measurement/non-zero indices or signal  &The network is adaptive to varying sparsity by controlling the number of LSTM units.\\\hline

CS + Data-Driven + Massive access &\cite{9390399}  &FNN &The measurement/sparse signal  &Authors combine the prior information from checking into network to enhance the performance.  \\\hline

CS + Data-Driven + Massive access &\cite{9252937}  &FNN &The measurement/sparse signal  &Authors extend the LAMP into multiple measurements and propose a new activation layer to utilize asynchronous information.  \\\hline

CS + Data-Driven + Massive access &\cite{9484069}  &FNN &The measurement/sparse signal  &The network has adaptive depth for varying sparsity.  \\\hline

CS + Data-Driven + Massive access &\cite{9508782}  &FNN &The measurement/sparse signal  &The network is obtained via unfolding the alternative direction method of multipliers algorithm.  \\\hline

CS + Data-Driven + Massive access &\cite{9508782}  &FNN &The measurement/sparse signal  &Authors reduce redundant parameters of the LISTA for efficiency.   \\\hline

CS + Data-Driven + Massive access &\cite{9174792}  &Autoencoder &The sparse signal/non-zero indices or signal &Authors propose to use autoencoder jointly learn the pilot matrix and receiver.  \\ \hline

\end{tabular}
}
\end{table*}

\textbf{Key assumptions}: The performance of activity detection and channel estimation assumes that the mMTC has a sporadic communication. This study is limited to contention-free access where each user has a unique pilot sequence and where the system is assumed to be synchronized.

\section{Data-Driven Compressed Sensing for mMTC}

\begin{figure*}[t]%
\centering%
\includegraphics[width=0.9\textwidth]{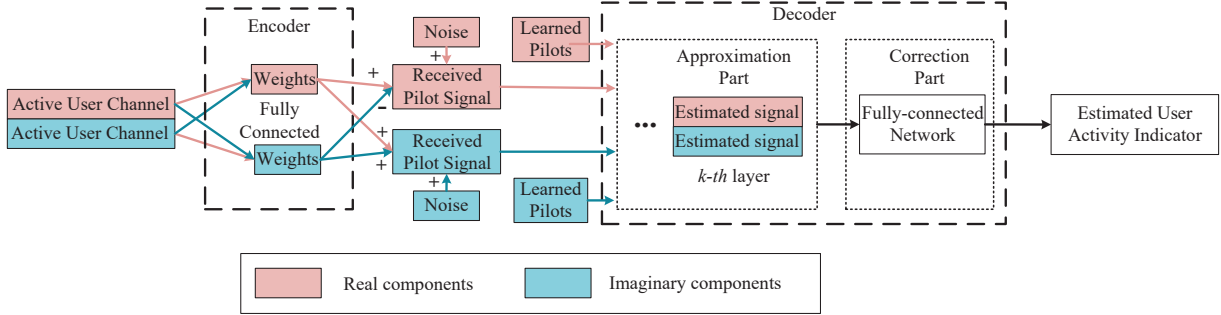}%
\DeclareGraphicsExtensions. \caption{An illustration of the network structure of the auto-encoder proposed in \cite{9174792} for joint pilot design and activity detection in grant-free access with MIMO.  } \label{fig:nn}
\end{figure*}%

\par
In this section, we first introduce different sparse problem formulations in communication systems and data-driven solutions. Table \ref{networks} summarizes several typical networks, outlining the problems they address in network design. We also show how to overcome the challenges of adapting standard NNs to the problem at hand.

\subsection{Problem formulation and data-driven CS solutions}

\par
As shown in Fig.~\ref{fig:CS-RA}, different scenarios lead to different sparse structures in the massive access problem. For a BS equipped with a single antenna, the massive access problem is detecting active users and estimating their channel with one measurement~\cite{9605579,9252937}. For a BS equipped with multiple antennas, we have multiple measurements and activity detection and data recovery is constructed as a row-sparse matrix recovery problem~\cite{9174792,9432908,9484069}. Furthermore, with some appropriate transform, the users' channel becomes sparse in the virtual angular domain in massive MIMO, especially in the mmWave domain. Those structures beyond the simple sparsity model, are desired to be exploited in the design of NNs for performance enhancement and parameter reduction.

\par
Data-driven methods have been applied for individual user activity detection~\cite{9432908,9634107}, joint user activity detection and channel estimation~\cite{9605579,9390399,9252937,9484069,9508782,9634107}, joint pilot design and activity detection~\cite{9174792}, and joint pilot design and activity detection and channel estimation~\cite{9174792}. Using data-driven methods for activity detection leads to a multi-label classification problem, where each label is a one-hot vector that corresponds to a distinct subset of active users. With estimated user activity, the BS can recover the channel coefficients of active users easily via least squares or minimum mean square estimator. In JADCE, data-driven methods directly learn the mapping between the received pilot signal and the product of user activity indicator and user channel vector. In the inference phase, the received signal at BS is fed into the trained network. The non-zero elements of the output of the network indicates the detected active users and their channel coefficients.

\par
Auto-encoders are used to design pilot sequences~\cite{9174792}. As shown in Fig.~\ref{fig:nn}, the input is the sparse channel vector, and the output is the user activity indicator or sparse channel vector. In the training phase, the encoder and the decoder are jointly trained. The pilot matrix design can be obtained according to the parameters in the encoder. The autoencoder can also be directly used for signal recovery. In the stacked denoising autoencoder in \cite{7447163}, the encoder is fixed as the pilot matrix and the stacked denoising autoencoder is equivalent to a general NN.


\subsection{Utilizing domain knowledge in communication systems}

\par
A promising direction in data-driven CS for mMTC is opened by utilizing various features of communication systems to design domain-specific NNs. In computer vision, the convolutional NN that considers the image structure information outperforms the fully-connected NN. However, most existing networks are not designed for communication systems. There exist various features of mMTC that can be utilized to enhance the NN design, i.e., the low-rank property in massive MIMO \cite{9432908}, the asynchronous information \cite{9390399}, various structure sparsity described in subsection A \cite{9484069}, etc. These communication-specific designs can be combined with various parts of NNs. For example, to use the row-sparse property in MIMO, we can use a modified activation function to enforce row-sparsity of estimated signal \cite{9484069}. Or we can design the network structure, such as constructing multiple networks in a distributed way, where each NN works on the detection at one antenna and various NNs can exchange the detected activity information to enhance the performance \cite{9390399}. By utilizing the special structure in different communication systems, those carefully designed networks show improved performance gain for activity detection and channel estimation \cite{9432908,9390399,9484069}.

\par
Another characteristic is considering the correlation embedded in the real and imaginary parts of the complex signal. As NNs cannot directly deal with complex signals, a common method is to flatten a complex signal into a real concatenated signal \cite{9484069,9508782,9605579}. Yet, this neglects the correlation between the imaginary and real parts and results in some performance degeneration. One method is to use different weights to process the real and imaginary parts of the complex signal separately, and combine the output to obtain the input of the next layer \cite{9174792}, as shown in Fig.~\ref{fig:nn}.

\par
In~\cite{9605579}, the checking mechanism is used to obtain the indices of successfully detected active users. This information is then used in the activity detection again, which corresponds to a sparse recovery problem with a known partial non-zero indices set. Those known non-zero indices are constructed as a one-hot vector that is used as additional input of the network and assist the selection of the remaining non-zero indices in the activation layer.

\begin{figure*}[t]%
\centering%
\includegraphics[width=0.9\textwidth]{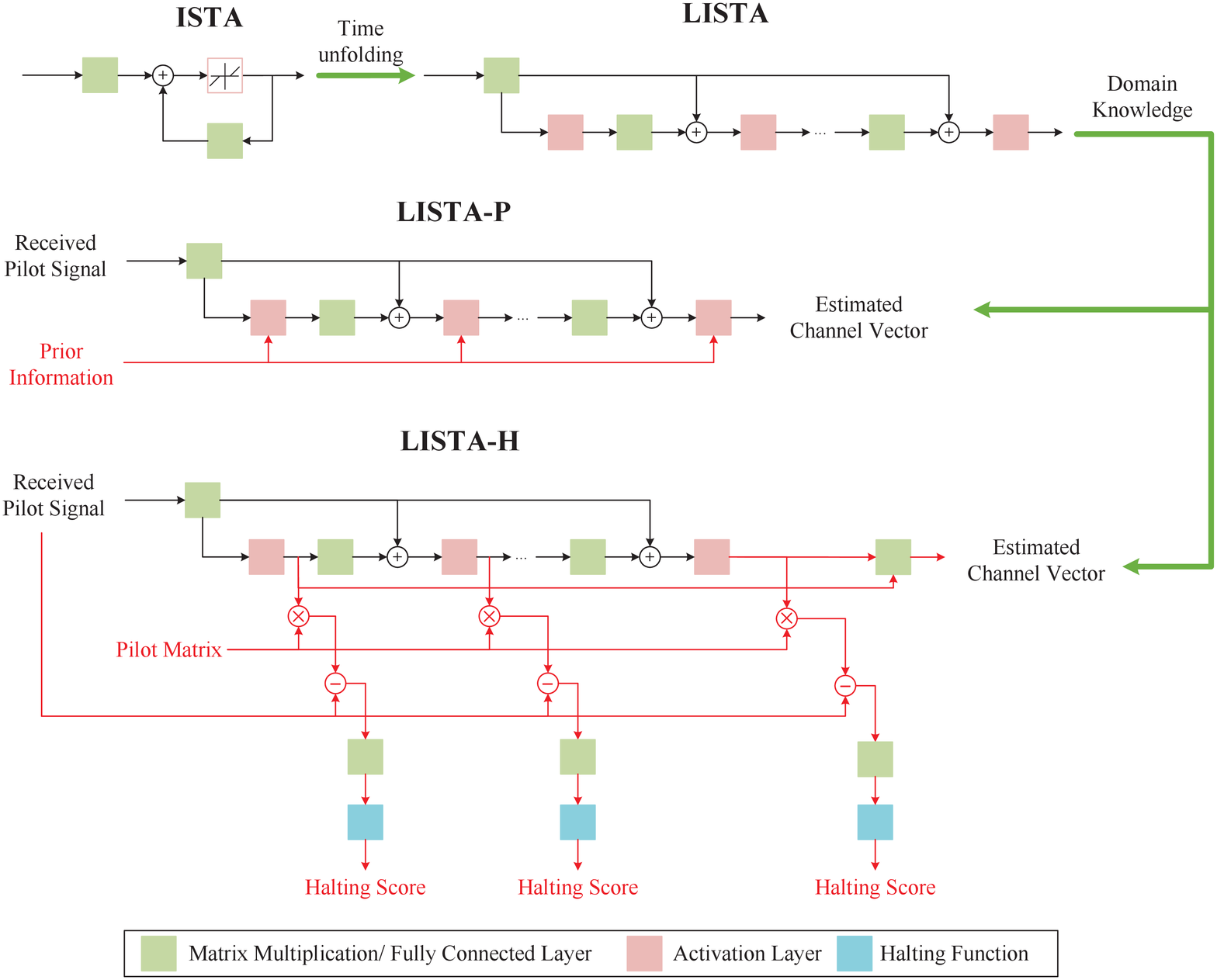}%
\DeclareGraphicsExtensions. \caption{Structures of deep-unfolding based networks, including LISTA \cite{gregor2010learning}, LISTA-P \cite{9605579}, and LISTA-H \cite{9252937}. } \label{fig:lista-h}
\end{figure*}%
\begin{figure}[!ht]%
\centering%
\includegraphics[width=0.38\textwidth]{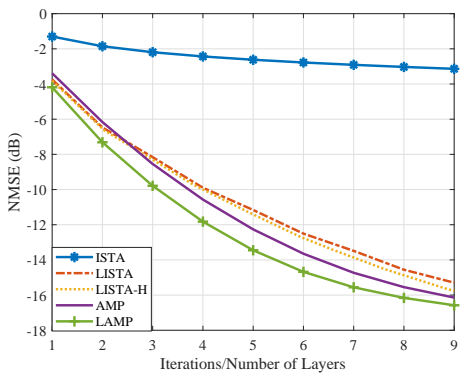}%
\DeclareGraphicsExtensions. \caption{The convergence rate and performance of the JADCE for the CS method and data-driven methods, respectively. The horizontal axis represents the number of iterations of ISTA and the number of layers of networks, which is related to the computational complexity. The vertical axis represents the normalized mean square error of the reconstructed sparse signal.} \label{fig:exp}
\end{figure}%

\begin{figure}[!ht]%
\centering%
\includegraphics[width=0.38\textwidth]{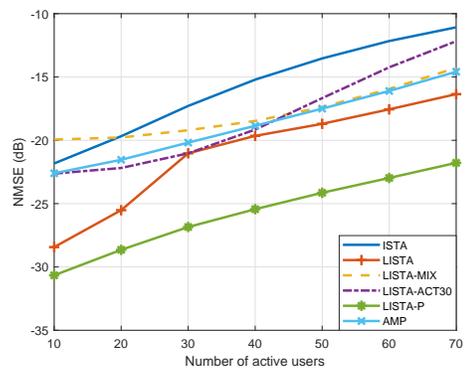}%
\DeclareGraphicsExtensions. \caption{Performance of the JADCE for the CS method and data-driven methods, respectively, under different number of active users.} \label{fig:exp1}
\end{figure}%

\subsection{Enhanced data-driven methods}

\par
Although above data-driven methods show outstanding performance, it also has drawbacks. A network trained with a specific pilot matrix cannot be applied to a different scenario associated with a distinct pilot matrix. The time-consuming training process hinders the application of data-driven methods.

To address the problem of retraining, model-agnostic meta learning (MAML) is used in~\cite{9500572} to speed up the training phase. While different situations with different pilot matrix can be considered as different tasks, MAML at first reuses the structural similarity across the scenarios to find a good initialization for the network in those tasks. Then, for a given pilot matrix, the network only needs a small number of iterations to adaptive to new task. In \cite{9601284}, Aberdam et al. directly combine the measurement matrix into the network and learn a universal architecture for different models. The measurement matrix and the measurement signal are jointly send to the learned network for recovery.

Another method is improving the training efficiency via decreasing the network parameters. For example, relying on rigorous analysis, Shi et al. reduce redundant the training parameters to simply the unfolding-based network \cite{9508782}. In \cite{9432908}, Shao et al. reduce the network parameters by utilizing the low-rank property of device state matrix that indicates the user activity and channel in MIMO, where the received pilot signal is mapped into a low-dimensional space and activity detection and channel estimation are performed in the low-dimension space.

\par
Another issue in data-driven methods is the fixed number of layers. While traditional algorithms for massive access can flexibly adjust the number of iterations for varying number of active users, the trained network in data-driven methods usually has a fixed number of layers. This directly influences the performance of activity detection in real applications, where the number of active users varies with time. In addition to the accuracy degradation, such a network is also inefficient in terms of computation as it has fixed access latency for different number of active users. To solve this problem, a novel network is proposed in~\cite{9252937}, using adaptive depths that can dynamically adjust the number of layers in executing specific tasks. In the proposed network, a mapping matrix is used to characterize the distribution of the reconstruction error of each layer, then a halting score is calculated to determine whether to stop at the current layer, and the weighted sum of the results of all layers is used as the output of the network. Another method to control the network depth is using the recurrent NN, i.e., LSTM networks. By controlling the number of LSTM cells in the network, Ahn et al. propose a network to support various scenarios with sparsity~\cite{9252937}. In each cell, the input and forget gates choose the active and inactive users respectively. Thus, the output of late stage traverses more cells and can thus estimate a signal with higher sparsity.

\subsection{Experimental results}

To show the performance gain of data-driven methods, we construct a simple CS-based grant-free access scenario as described in Section II. We assume $500$ users and each has a unique pilot sequence of the length $250$. Elements in the pilot sequences are randomly generated following the independent and identically distributed Gaussian distribution. In practice, some finite alphabet symbols are used as pilots. The probability of activation is set to $0.1$ and is identical for all devices. The signal-noise-ratio is set to $20$ dB. We compare the normalized mean square error (NMSE) of ISTA, LISTA \cite{gregor2010learning}, LISTA-H which has an adaptive number of layers~\cite{9252937}, AMP, and LAMP \cite{7934066}.

As shown Fig.~\ref{fig:lista-h}, the ISTA only contain matrix multiplication and thresholding function. By unfolding the ISTA, we can obtain the LISTA, where the matrix multiplication is realized via the fully connected layer, and the activation function is the soft-thresholding function with learned thresholding parameters. Compared with LISTA, LISTA-P adds the prior knowledge related to known non-zero indices sets to the activation function to help the sparse supports selection. LISTA-H adds a halting function in each layer to evaluate whether to stop at this layer by comparing the received pilot signal with the product of the output at this layer and the pilot matrix. LAMP is the unfolding-based network of AMP.

To compare the computation complexity of ISTA, AMP and LISTA, we evaluate them on the same test data set, using a computer with a quad-core 2.3GHz CPU and 16 GB RAM. The AMP costs 1.06s (30 iterations) and its NMSE reaches -17.45dB. The NMSE performance of ISTA is -17.49dB which costs 40.9s (1000 iterations). The LISTA, whose NMSE reaches -18.31dB with 16 layers, only uses 0.24s.

In Fig.~\ref{fig:exp}, we can see that the LISTA and LAMP converge faster and perform better than ISTA and AMP. This is because the original activity detection and channel estimation is an NP-hard problem and the ISTA solves a convex relaxed optimization problem, which leads to reconstruction errors when the number of measurements is smaller than some threshold. Data-driven methods directly learn the mapping between the received pilot signal and channel of active users and provides the possibility to learn the optimal estimation result. As LISTA-H can flexibly adjust the number of layers when varying the sparsity, it has a lower average number of layers.

We also measure the training time of networks with different depths. We observe the network with more layers requires more training time. For example, in comparison to a network with 10 layers, the network with 20 layers consumes $47.77\%$ more training time and its NMSE reduces $13.71\%$, which illustrates the trade-off between computation/training time and performance.

In Fig.~\ref{fig:exp1}, we compare the performance under different numbers of active users. LISTA-ACT30 is trained under the sparsity level 30, thus has the best test performance on the sparsity level 30 and decreased performance on other sparsity levels. The LISTA-MIX, which is trained under mixed sparsity levels, is more robust than LISTA-ACT30. The LISTA is trained for each sparsity level, thus has better performance. Yet, training networks for all settings is time-consuming. Training a network with data of mixed settings would be less time-consuming, while its performance is usually limited. Compared with LISTA, LISTA-P~\cite{9605579} utilizes the error-checking information as the prior information of the network and achieves outstanding performance.

\section{Challenges and Future Directions}

Data-driven methods are indispensable tools for 6G systems, but there are still significant challenges for efficient usage in various applications. In this section, we present the main challenges and possible research directions for data-driven CS in mMTC.

\subsection{Constructing the training data}

A key issue in the data-driven method for massive access is where does the training data come from. Most existing research on data-driven methods for mMTC mainly use generated data from simulators according to some channel model. Nevertheless, the existing model-driven access schemes fail to consider complex cell-specific channel characteristics in mMTC. With data-driven methods, it is possible to learn more scenario-based characteristics by exploiting the historical channel state information stored in BS. In situations where the historical data is not available, one can rely on an unmanned aerial vehicle to scan the whole scenario and construct a three-dimensional map, and then use techniques such as ray tracing to generate channel data. In addition to the channel information, the training data may contain more information, such as the underlying traffic pattern of users. According to the historical access data, data-driven methods can learn the access probability of users and predict frequently activated users with a higher probability. If only a small part of historical data is available, we can train the network with simulated data and fine-tune the trained model on the limited real data. In practice, the trained model can be frequently updated by using the newly available data. Another promising venue is the creation of synthetic data using generative models, such as variational autoencoder and generative adversarial networks.

\subsection{Designing specialized networks}
Most well-known NNs in data-driven methods are originally designed for processing visual signals and natural language. To fully exploit the potential of data-driven methods, it is therefore necessary to study dedicated NNs for communication systems, for example, the NNs working with complex signals. Besides, in massive access, there are a lot of features that can be taken into account in the NN design, such as the user activity patterns. We can use some recurrent NN to capture the temporal correlation of user activity. In recent years, supporting mMTC with massive MIMO is attractive. However, this setup increases significantly the signal dimension, which puts forward a training challenge for high-dimensional networks. A possible method to deal with the prohibitive computational complexity due to large-scale antennas is designing light networks to reduce the number of parameters of the network, e.g., by exploring the low-rank property in massive MIMO~\cite{9432908}. Besides, in 6G, massive MIMO further extends to cell-free massive MIMO, which has different characteristics with massive MIMO and urges for improvements in access system design. In cell-free massive MIMO, the channel estimation is more difficult as signals at different antennas have different large-scale fading. Specific features, such as the spatial sparsity of signals, can inspire additional network optimization. For instance, we can use a new sparsity-enforcing activation function in the original sparse recovery network in order to attain the spatial sparsity.

\subsection{Optimizing the training process}

Data-driven methods in the inference phase show improved computation complexity in various literature, which is important for latency-sensitive applications such as real-time monitoring. Yet, this comparison happens in the inference phase and the time-consuming training phase is still a considerable issue. Specifically, in most cases, we have to re-train a new network if some settings in the scenario are changed, e.g., the number of access users. Thus, optimizing the training process of data-driven methods to fit the flexible applications in 6G is a challenging problem. A promising direction is seen in meta-learning \cite{9500572}, as it enables the network to learn a new task quickly. To obtain a good meta-learning model with powerful adaptive and generalization capabilities, we need to train the network under different access scenarios, i.e., different pilot matrix. Then, for a new scenario with new pilot matrix, the network can learn fast, using only a few access data. Thus, we expect meta-learning to amend the deficiency of the existing data-driven methods, especially when only a limited amount of historical access data is available for training.

\subsection{Applying the trained network}

Data-driven methods for massive access have shown outstanding performance in plenty of literature, but we still have a long way to go to use the data-driven methods in practice. The massive access scenarios in practical applications are often more complex than the models in the literature, e.g., dynamic access densities, different types of devices, different sizes of data packets, and so on. The practical scenario with dynamic user population can be addressed by contention-based CS access, where active users randomly select the pilots from the common pilot pool. Then, we only need to train one network for each pilot matrix.

The diversity between the distribution of the network input and training data results in performance degradation. In existing works, the network trained under certain configurations, e.g., fixed signal-to-noise ratio or fixed active probability, has obvious performance degradation under different settings. A mixed training data set with different settings may help to improve the robustness of data-driven methods. Yet, the limited learning capability of the network also result in performance degradation, while networks with more parameters face the problem of increased training time and computation complexity issues. Alternatively, we can train multiple networks for different SNRs, and choose the suitable one at the time of inference time. Or we can use transfer learning to quickly fit the trained network to a specific SNR. Another direction is integrating the SNR into network design, like the network in \cite{9252937}.

\section{Conclusions}

In this paper, we discuss the data-driven CS for grant-free random access, which performs the activity detection and channel estimation via using a NN to solve the sparse recovery problem. Data-driven methods decrease the computational complexity of the receiver and enhance performance, mainly due to leveraging domain knowledge from communication systems into the NN design. We show how the data-driven methods can improve the performances, and show several variations of NNs for CS-based grant-free random access. Finally, we summarize the challenges and future directions of data-driven CS methods for mMTC.

\ifCLASSOPTIONcaptionsoff
  \newpage
\fi



%
%
%

\bibliographystyle{IEEEtran}
\bibliography{refs}

%



\begin{IEEEbiographynophoto}{Yanna Bai}
(yannabai@bjtu.edu.cn) received both the B.Eng. and M.Eng. degrees from Beijing Jiaotong University in 2016 and 2018, respectively, and is pursuing the Ph.D. degree with the same university. Her current research interests include AI-assisted communications and massive machine-type communications.
\end{IEEEbiographynophoto}


\begin{IEEEbiographynophoto}{Wei Chen}
(weich@bjtu.edu.cn) (M'13-SM'18) received the B.Eng. degree and M.Eng. degree from Beijing University of Posts and Telecommunications, China, in 2006 and 2009, respectively, and the Ph.D. degree in Computer Science from the University of Cambridge, UK, in 2013. Later, he was a Research Associate with the Computer Laboratory, University of Cambridge from 2013 to 2016. He is currently a Professor with Beijing Jiaotong University, Beijing, China. His current research interests include intelligent, wireless communication systems and multimedia processing.
\end{IEEEbiographynophoto}

\begin{IEEEbiographynophoto}{Feifei Sun}
(feifei.sun@samsung.com) received her Ph.D. in Communication and Information Engineering from the Beijing University of Post and Telecommunication, China. She is now a principal engineer in Samsung Research China - Beijing (SRC-B). Her research interests in B5G/6G wireless communications standardization on IoT related techniques such as NB-IoT, eMTC, RedCap, URLLC, as well as AI/ML in physical layer.
\end{IEEEbiographynophoto}

\begin{IEEEbiographynophoto}{Bo Ai}
(boai@bjtu.edu.cn) (M'00-SM'10) received the M.S. and Ph.D.degrees from Xidian University, Xian, China, in 2002 and 2004, respectively. He was with Tsinghua University, Beijing, China, where he was an Excellent Postdoctoral Research Fellow in 2007. He is currently a Professor and an Advisor of Ph.D.candidates with Beijing Jiaotong University, Beijing, where he is also the Deputy Director of the State Key Laboratory of Rail Traffic Control and Safety. He is also currently with the Engineering College, Armed Police Force, Xian. His interests include the research and applications of OFDM, high-power amplifier linearization techniques, radio propagation and channel modeling, global systems for mobile communications for railway systems, and LTE for railway systems.

\end{IEEEbiographynophoto}

\begin{IEEEbiographynophoto}{Petar Popovski}
(petarp@es.aau.dk) is a Professor at Aalborg University and a Visiting Excellence Chair at University of Bremen. He received Dipl.-Ing. (1997)/Mag.-Ing. (2000) in communication engineering from Sts. Cyril and Methodius University in Skopje and Ph. D. from Aalborg University (2004). He is a Fellow of IEEE, holder of an ERC Consolidator Grant (2015-2020), Villum Investigator, and a Member at Large on the Board of Governors in IEEE Communication Society. His research interests are in wireless communications/networks and communication theory. He authored the book "Wireless Connectivity: An Intuitive and Fundamental Guide", published by Wiley in 2020.
\end{IEEEbiographynophoto}






\end{document}